\documentclass[aps,twocolumn,superscriptaddress]{revtex4}
\usepackage{graphicx}
\usepackage{rotating}
\usepackage{amssymb}
\usepackage{psfig}

\begin{document}

\title{Poincar\'e recurrence and measure of hyperbolic and nonhyperbolic 
chaotic attractors}
\def\active{0} 

\author{Murilo S. Baptista}
\affiliation{Universit{\"a}t Potsdam, Institut f{\"u}r Physik, Am Neuen Palais
10, D-14469 Potsdam, Germany} 
\author{Suso Kraut} 
\affiliation{Instituto de F\'isica, Universidade de S\~ao Paulo, Caixa Postal
66318, 05315-970 S\~ao  Paulo, Brazil} 
\author{Celso Grebogi} 
\affiliation{Instituto de F\'isica, Universidade de S\~ao Paulo, Caixa Postal
66318, 05315-970 S\~ao  Paulo, Brazil} 
\date{\today}

\begin{abstract}
We study Poincar\'e recurrence of chaotic attractors for regions of finite
size. Contrary to the standard case, where the size of the recurrent regions
tends to zero, the measure is not supported anymore solely by unstable 
periodic orbits inside it, but also by other special recurrent trajectories, 
located outside that region. The presence of the latter leads to a deviation 
of the distribution of the Poincar\'e first return times from a Poissonian. 
Consequently, by taken into account the contribution of these recurrent 
trajectories, a corrected estimate of the measure can be provided. This has 
wide experimental implications, as in the laboratory all returns can 
exclusively be observed for regions of finite size only.  
  


\end{abstract}

\maketitle

In dynamical systems one is often interested in calculating asymptotic 
invariant physical quantities that are independent of the initial conditions
and invariant under the evolution of the dynamics. Those invariant quantities 
are the main subject of ergodic theory, which was suitably applied to chaotic 
dynamical system, producing important average estimates, like Lyapunov 
exponents and dimensions \cite{eckmann:1985}. Chaotic systems are ergodic and
therefore time averaged quantities calculated from a typical trajectory can
also be calculated through space integrals using the natural measure of a
chaotic attractor, which is related to the probability of finding a typical 
trajectory in some region of the attractor. Hence, the calculation of this
measure yields essential observable quantities of chaotic systems. 
\\
\indent
It is well accepted that the support of the measure as well as properties of 
the attracting set are hierarchically approached by the set of unstable
periodic orbits (UPOs) which are embedded in the chaotic attractor
\cite{auerbach:1987,UPO,biham:1989}. For hyperbolic systems this approach 
allows a reconstruction of the fractal dimension of a chaotic attractor 
through UPOs \cite{auerbach:1987,grebogi:1988}. For the important class of 
nonhyperbolic chaotic attractors, i. e., when there exist UPOs whose stable 
and unstable manifolds exhibit tangencies, only numerical results for averages
in phase space exist, suggesting that, on average, the nonhyperbolic regions
do not play a special role \cite{lai:1997}. 
\\
\indent
Recently, an exact result on the probability distribution of the series of 
first return times (FRTs) $\tau_i, i=1,\ldots,N$ for measurable dynamical 
systems (including nonhyperbolic ones) was provided \cite{hirata:1999}. It 
was proven that the distribution of the FRTs, if the return time is larger 
than some constant value and the size of the region goes to zero, approaches 
the Poissonian $\rho(\tau,\mathcal{B})=\mu\exp{^{-\mu\tau}}$, with  $\mu$ 
being the probability density measure of the region $\mathcal{B}$ of size 
$\epsilon$, as also discussed in \cite{zaslavsky:1991,baptista:2000}. Thus, 
\begin{equation}
\lim_{\epsilon \rightarrow 0}\rho(\tau>\frac{\mu}{C},\mathcal{B})     -      
\mu\exp{^{-\mu\tau}}     < G(\mu(\mathcal{B}))
\label{limit}
\end{equation}
\noindent
with  $C$ being a suitable normalizing factor and $G(\mu(\mathcal{B}))$ a 
function of the probability density measure, $\mu$, in $\mathcal{B}$.
The multifractal spectrum of return times has been investigated as well
\cite{hadyn:2002}.
\\
\indent
In this work, we extend this rigorous treatment to the case for which 
$\epsilon$ is far from zero and $\tau$ is arbitrary. This is motivated by 
the fact that experimentally only returns to regions of finite size can be 
observed. In the following we take the returnung region to be a square of
size $\epsilon$ intersecting the invariant set (attractor). 
We show that for both, nonhyperbolic (logistic map and H\'enon map)
and hyperbolic (Cat map) systems, the presence of recurrent trajectories that 
are not associated with UPOs inside the region increases as one increases 
the size of the region, causing the measure to become abnormally singular. 
In the case of nonhyperbolic systems, even for square of very small but 
finite size, the contribution to the measure of these recurrent trajectories 
is already large if the interval is close to a homoclinic tangency. Hence, 
as one observes a dynamical system for intervals of finite size, the local 
dynamics in $\mathcal{B}$ can neither be completely governed by the 
linearization of these UPOs as proposed in the theorem of Hartman-Grobman 
\cite{hartman:1960}, nor the measure inside the interval, $\mu(\mathcal{B})$, 
can be exactly calculated through the UPOs inside it. However, from the 
distribution of the FRTs we show that it is possible to calculate the 
measure exactly by introducing a correction of the measure from the 
eigenvalues of the UPOs inside the interval, when the intervals are far 
from small.
\\
\indent
As already mentioned, an important result concerns the reconstruction of the 
measure using UPOs \cite{grebogi:1988}. For hyperbolic systems, one can 
derive a formula to calculate the probability density, $\mu(\mathcal{B})_
{EIG}$ of small squared subregions $\mathcal{B}$ in the attractor. More
explicitely 
\begin{equation}
\mu(\mathcal{B})_{EIG}=\sum_{k=1}^{N_j}{\left(
\frac{1}{L_k}\right)}
\label{mu_EIG}
\end{equation}
\noindent
where the $L_k$ are the positive eigenvalues of all fixed points located in 
$\mathcal{B}$ of the $j$-fold iterate of the map, (i. e., the fixed points 
are period-j UPOs $\in \mathcal{B}$), and $N_j$ represents the number of 
period-$j$ UPOs $\in \mathcal{B}$. 
\\
\indent
We conveniently define a finite sized region to have a hyperbolic character 
if that region exhibits a distribution of FRTs close to a Poissonian. 
In this case, the measure, as well as the dynamics in that region, is 
described by the UPOs inside it, and typically
there lie neither homoclinic tangencies nor low-period UPOs inside the region.
Analogously, we define that a region is nonhyperbolic if the distribution of 
the FRTs deviates from a Poisson law.
In that case, recurrent trajectories play an important role in the dynamics 
and in the contribution to the measure; here, typically low-period UPOs or 
homoclinic tangencies are found. We show that the larger the contribution of 
the recurrent trajectories to the measure is, the larger is the deviation of 
the calculation of the measure from the eigenvalues of the UPOs inside these 
intervals. The contribution becomes negligible as either the interval size
tends to zero or the intervals are placed in hyperbolic regions and off 
homoclinic tangencies and low-period UPOs. 
\\
\indent
The probability measure in a square $\mathcal{B}$ due to all the recurrent 
orbits (including the UPOs) with recurrent periods between $\tau_k$ and 
$\tau_l$ can be calculated from the distribution of the FRTs 
$\rho(\tau_i,\mathcal{B})$ as 
\begin{equation}
\mu(\tau_k, \tau_l) = 
\frac{1}{\langle \tau (\mathcal{B}) \rangle}\sum_{i=k}^{l} 
\rho(\tau_i,\mathcal{B}),
\label{measure_discrete}
\end{equation}
\noindent
where the average first return time is given by 
\begin{equation}
{\langle \tau(\mathcal{B}) \rangle} = \lim_{n \rightarrow \infty} \frac{1}{n}\sum_{i=1}^n \tau_i(\mathcal{B}).
\label{average_return} 
\end{equation}
Note that if one sets in Eq. (\ref{measure_discrete}) $\tau_k=\tau_{min}$ 
(minimal FRT in $\mathcal{B}$) and $\tau_l=\tau_{max}$ (maximal FRT in 
$\mathcal{B}$), then, $\sum_{i=\tau_{min}}^{\tau_{max}} \rho(\tau_i,
\mathcal{B}) = 1$ and one recoveres Kac's theorem, which
states that
\begin{equation}
\mu(\mathcal{B})=\frac{1}{\langle \tau(\mathcal{B}) \rangle},
\label{kac_lemma}
\end{equation}
\noindent
with $\langle \tau(\mathcal{B}) \rangle$ the average FRT, relating a time
averaged quantity  $\langle  \tau(\mathcal{B}) \rangle$ with a spatial
quantity, $\mu(\mathcal{B})$.  
\\
\indent
In order to calculate the measure exclusively due to UPOs inside 
$\mathcal{B}$, we hypothezise that the distribution of the FRTs can be split 
into two discrete functions (for a small FRTs) and into two continuous 
functions (for large FRTs), one describing the contribution due to the UPOs
and the other the contribution of the  recurrent trajectories,
respectively. Hence,  
\begin{equation}
\mu^{\prime}(\mathcal{B}) = \mu(\mathcal{B})_{REC}^{d} + \mu(\mathcal{B})_{UPO}^{d} + \mu(\mathcal{B})_{REC}^c + \mu(\mathcal{B})_{UPO}^c,
\label{break_summation}
\end{equation}
\noindent
with $\mu(\mathcal{B})_{REC}^d = \frac{1}{\langle \tau (\mathcal{B})\rangle} 
\sum_{i=\tau_{REC}} \rho(\tau_i,\mathcal{B}) \label{mu_rec_disc}$, 
$\mu(\mathcal{B})_{UPO}^d = \frac{1}{\langle \tau (\mathcal{B})\rangle} 
\sum_{i=\tau_{UPO}} \rho(\tau_i,\mathcal{B}) \label{mu_upo_disc}$, and 
\begin{eqnarray}
\mu(\mathcal{B})_{REC}^c  &=& \frac{1}{\langle \tau (\mathcal{B}) \rangle} 
\int_{\tau^{min}_{UPO}}^{\tau_{max}} \beta_{REC} \exp^{(-\alpha \tau)} 
\label{mu_rec_cont} d\tau\\
\mu(\mathcal{B})_{UPO}^c  &=& \frac{1}{\langle \tau (\mathcal{B}) \rangle}
\int_{\tau^{min}_{UPO}}^{\tau_{max}} \alpha \exp^{(-\alpha \tau)} d\tau.
\label{mu_upo_cont}
\end{eqnarray}
\noindent
The index $d$ designates a discrete summation and $c$ indicates a 
continuous integral. $\mu(\mathcal{B})_{REC}^d$ is the measure due to short
recurrent trajectories, $\mu(\mathcal{B})_{UPO}^d$ due to low-period UPOs, 
$\mu(\mathcal{B})_{REC}^c$ 
the measure due to long recurrent trajectories and $\mu(\mathcal{B})_{UPO}^c$
due to high-period UPOs \cite{periodlength}. The return time $\tau_{REC}$ 
stands for first returns that are different from any of the period of the UPOs
inside $\mathcal{B}$, and  $\tau_{UPO}$ for all periods of the UPOs.
The time  $\tau^{min}_{UPO}$ corresponds to the minimum period for which 
one can consider the distribution of FRT to be well described by a continuous 
exponential function and the measure can be calculated with the Eqs. 
(\ref{mu_rec_cont}) and (\ref{mu_upo_cont}). These two integrals are 
constructed under the assumption that the probability distribution is given by 
\begin{equation}
\rho(\tau,\mathcal{B})=\beta\exp{^{-\alpha\tau}},
\label{return_dist}
\end{equation}
i.e., it agrees with Eq. (\ref{limit}). Since $\mu(\mathcal{B}) \propto 
\epsilon^{D_p}$, $D_p>0$ being the pointwise dimension for that interval 
\cite{farmer:1983}, as $\epsilon \rightarrow 0$ and $\tau>\frac{\mu}{C}$,  
the distribution (\ref{return_dist}) should  approach a Poisson function 
$\rho(\tau,\mathcal{B})=\mu\exp{^{-\mu\tau}}$. Equation (\ref{return_dist})
can be broken in a summation of two distributions, one due to the long 
recurrent trajectories $\rho_{REC}^c = \beta_{REC} \exp^{(-\alpha \tau)}$, 
and the other due to the large-period UPOs, a Poissonian of the type 
$\rho_{UPO}^c = \alpha \exp^{(-\alpha \tau)}$, such that 
$\rho_{REC}^c + \rho_{UPO}^c$ = $\beta \exp^{(-\alpha \tau)}$. Therefore, 
$\beta_{REC} + \alpha$ = $\beta$, and, from Eq. (\ref{limit}), as $\epsilon 
\rightarrow 0$,  $\beta_{REC} =0$, and $\alpha=\mu$.  
The coefficients $\beta$ and $\alpha$ are obtained by fitting the 
distribution of the FRTs by an exponential of the form of Eq. 
(\ref{return_dist}). Integrating Eqs. (\ref{mu_rec_cont}) and 
(\ref{mu_upo_cont}) we get $\mu_{REC}^c$ = $\left( \frac{\beta}{\alpha}-1 
\right) \gamma \mu$ and $\mu_{UPO}^c$ = $\gamma \mu$, with $\gamma$ = 
$\exp^{- (\tau^{min}_{UPO}) \alpha} - \exp^{- (\tau^{max}) \alpha}$. 
\\
\indent
Assuming that $\mu_{REC}^d$ and  $\mu_{UPO}^d$ are negligible in comparison 
with $\mu_{REC}^c$ and $\mu_{UPO}^c$, and that $\mu_{EIG}$ = $\mu_{UPO}^c$ 
(what is true for moderately nonhyperbolic regions which are predominant 
in the attractor), we arrive at a formula that expresses the value of the 
measure through the measure calculated exclusively with the contribution 
from the UPOs inside $\mathcal{B}$
\begin{equation}
\mu(\mathcal{B}) \cong K \mu(\mathcal{B})_{EIG},
\label{correction}
\end{equation}
\noindent
where $\frac{1}{K}$ = $\left( 1- \gamma [\frac{\beta}{\alpha} - 1] \right)$.
\\
\indent
Our model of a nonhyperbolic system is the H\'enon map, 
$H:\mathcal{X} \rightarrow \mathcal{X}$, $x_{i+1}=a-x_i^2+b\,y_i$, and 
$y_{i+1}= x_i$, with $a$ = $1.4$ and $b$ = $0.3$. The nonhyperbolicity of  
this map is due to tangencies of the stable and unstable manifolds of 
periodic orbits embedded in the chaotic attractor.  
\\
\indent
In Fig. \ref{rec_fig1} we show what should be expected from a hyperbolic 
and from a nonhyperbolic region. In Fig. \ref{rec_fig1}(a), points 
represent all the UPOs up to period 23, calculated using the method of 
Ref. \cite{biham:1989}, and stars depict the primary tangencies together 
with their five images and preimages, where primary indicates that the 
curvature of the manifold is minimal in their neighborhood.
In Fig. \ref{rec_fig1}(b) we display a typical nonhyperbolic region (box) 
centered at the primary tangency $T_1: (x,y) = (1.7801,-0.0949)$ with box size 
$\epsilon=0.02$. The lowest periodic orbit found in that region has period  
18,  whereas the smallest FRT is $\tau$ = 9. Points represent the H\'enon 
attractor, and the filled circle a component of a period-9 orbit. The stripe 
marked by  $W_s$ pictures points inside $\mathcal{B}$ that under $H^9$ (the
9-fold iterate of the map $H$) remain in the interval, represented by $W_u$. 
As one can see, the stripes $W_s$ and $W_u$ are aligned along the stable and 
unstable manifold of the period-9 orbit located outside $\mathcal{B}$. 
Consequently, the tangency creates recurrent trajectories that are not   
associated with any UPO inside $\mathcal{B}$. Note that, as could be 
anticipated for a nonhyperbolic region, the stripes $W_s$ and $W_u$ are almost
parallel close to $T_1$. In Fig. \ref{rec_fig1}(c) we demonstrate that  
a similar effect occurs when there is a low-period UPO inside $\mathcal{B}$
(here of period-2). Although there exists no period-13 UPO, we nevertheless
find a FRT of $\tau$=13. That is caused by points on the stripes $W_s$ and 
$W_u$, which lie on the stable and unstable manifold of a period-13 orbit
outside $\mathcal{B}$.  Finally, we show in Fig. \ref{rec_fig1}(d) a 
typical hyperbolic region, where to all observed FRTs of length $\tau$ 
a period-$\tau$ UPO inside $\mathcal{B}$ corresponds, and the manifolds  
cross transversally. 
\begin{figure}[!h]
\caption{Points represent the chaotic attractor  $\Gamma$, filled circles 
periodic orbits, and $W_s$ denotes points inside $\mathcal{B}$ that under 
$H^P$ (where $P$ is the period of the UPO) remain in the interval; their
$P$-fold iteration is shown by $W_u$. In (b) the box is centered in a 
primary tangency and in (c) in a low-period UPO. Both are typical 
nonhyperbolic regions. In (d) a characteristic hyperbolic region is depicted.}
\label{rec_fig1}
\end{figure}
\\
\indent
To quantify now these findings with our theory, we analyze two specific 
intervals with $\epsilon=0.02$, a nonhyperbolic one, $\mathcal{B}_1$, centered
at the primary tangency $T_1$, see Fig. \ref{rec_fig1}(b), and a hyperbolic 
one, $\mathcal{B}_2$, centered at the point $(x,y)=(1.1181,0.1472)$, see Fig. 
\ref{rec_fig1}(d), for which the smallest period of all UPOs is found to be
21, and no tangency is present. 
\\
\indent
For the nonhyperbolic interval $\mathcal{B}_1$, we get $\mu_{REC}^d = 1.28 
\times 10^{-4}$ from the return times $\tau_{REC}=(9,16,17,18,19,20,22)$,
$\mu_{UPO}^d$=0, $\mu_{REC}^c = 2.41 \times 10^{-5}$, and $\mu_{UPO}^c = 2.885
\times 10^{-3}$, with $\tau^{min}_{UPO} = 23$ and $\tau_{max} = 4284$ in Eqs. 
(\ref{mu_rec_cont}), (\ref{mu_upo_cont}), obtained by measuring 400,000 
returns to that interval. Employing Eq. (\ref{break_summation}), this yields 
a measure of $\mu^{\prime} = 3.0923 \times 10^{-3}$. We stress that the
measure evaluated using Kac's lemma with 400,000 returns,  considerered to be
the exact one, leads to $\mu =  3.100 \times  10^{-3}$, very close to our
estimate from Eq. (\ref{break_summation}). We see  further that most of the
measure is due to the UPOs and the contribution from recurent trajectories
$\mu_{REC}^d$ and $\mu_{REC}^c$ can be neglegted. From Eq. (\ref{mu_EIG}) the
best estimate (determined for UPOs of up to period 30) is $\mu_{EIG} = 3.003
\times 10^{-3}$. Note that $\mu_{EIG} \cong \mu_{BOTH}^d + \mu_{UPO}^c$. In
spite of this region being strongly nonhyperbolic, we apply nevertheless the
correction in Eq. (\ref{correction}), which  yields  $\mu(\mathcal{B}) =
3.027 \times 10^{-3}$, a value closer to the exact $\mu$ than $\mu_{EIG}$.  
\\
\indent 
The hyperbolic region $\mathcal{B}_2$ results in $\mu_{REC}^d = 0$, 
$\mu_{UPO}^d = 0$, $\mu_{REC}^c = 9.97 \times 10^{-6}$, and 
$\mu_{UPO}^c = 7.712 \times 10^{-4}$, with $\tau^{min}_{UPO} = 23$ and 
$\tau_{max} = 16871$ obtained for 400,000 returns to that interval as before.
This gives with Eq. (\ref{break_summation}) $\mu^{\prime} = 7.2295 \times
10^{-4}$, whereas the exact measure, using Eq. (\ref{kac_lemma}), ensues $\mu
= 7.2422 \times 10^{-4}$, very close to the value obtained with the correction
formula Eq. (\ref{break_summation}). Again, most of the measure is due to the
UPOs. From Eq. (\ref{mu_EIG}) the best estimate (calculated for UPOs of up to
period 30) of $\mu_{EIG} = 7.5312 \times 10^{-4}$. For this case,
Eq. (\ref{correction}) has no use since $\mu_{EIG} > \mu$. 
\begin{figure}[!h]
\includegraphics[angle=0,width=7cm,height=5cm]{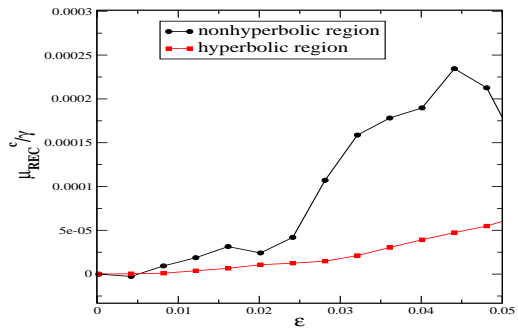}
\caption{The weighted measure $\frac{\mu(\mathcal{B}_i)_{REC}^c}{\gamma}$,
i=1, 2, due to recurrent trajectorie with respect to $\epsilon$ for the
interval $\mathcal{B}_1$ (circles) and $\mathcal{B}_2$ (squares).} 
\label{hip_nhip_fig2}
\end{figure}
\\
\indent
To understand how the recurrent trajectories contribute to the measure of
these two intervals when the boxsize $\epsilon$ varies, in Fig. 
\ref{hip_nhip_fig2} the value of $\frac{\mu(\mathcal{B}_i)_{REC}^c}{\gamma}=
(\frac{\beta} {\alpha}-1) \mu(\mathcal{B}_i), i=1, 2$ is plotted against
$\epsilon$. It can be calculated using only the information of the FRTs, and
thus there is no need of knowing $\tau^{min}_{UPO}$, a numerically envolving
task. We see that for the interval $\mathcal{B}_1$, centered at the primary
tangency $T_1$, the contribution of the recurrent trajectories
$\mu_{REC}(\mathcal{B}_1)$ is much larger than $\mu_{REC}(\mathcal{B}_2)$ for
the interval $\mathcal{B}_2$, provided $\epsilon>0.0002$. It grows moreover
much faster with increasing $\epsilon$. Also, for $\mathcal{B}_2$ this
contribution decays smoothly as one decreases $\epsilon$. This smooth decay in
arbitrary intervals of finite size takes typically place in hyperbolic regions
when the size appoaches zero. It is also encountered in the hyperbolic cat map
as well as in the hyperbolic regions of the logistic map \cite{baptista:2000}. 
\begin{figure}[htb]
\includegraphics[angle=0,width=7cm,height=6cm]{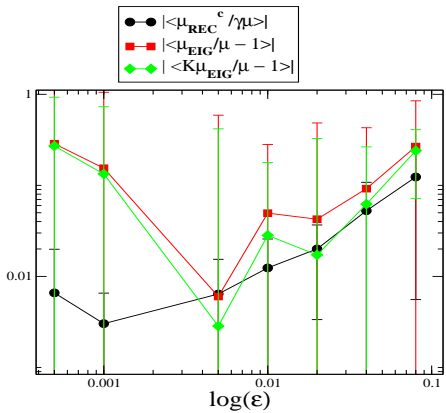}
\caption{Averages $|\langle (\frac{\mu(\mathcal{B}_i)_{REC}^c}{\gamma\mu
(\mathcal{B}_i)}) \rangle|$ (squares), $|\langle (\frac{\mu(\mathcal{B}_i) 
- \mu(\mathcal{B}_i)_{EIG}}{\mu(\mathcal{B}_i)} \rangle|$ (circles), and 
$|\langle (\frac{ \mu(\mathcal{B}_i) - K \mu(\mathcal{B}_i)_{EIG}}
{\mu(\mathcal{B}_i)} \rangle|$ (diamonds), $i=1,2,...,5000$, against 
$\epsilon$, in a log-log graph.}
\label{hip_nhip_fig3}
\end{figure}
\\
\indent
In Fig. \ref{hip_nhip_fig3} we plot $|\langle (\frac{\mu(\mathcal{B}_i)_{REC}
^c}{\gamma\mu(\mathcal{B}_i)}) \rangle|$ (squares), $|\langle (\frac{\mu
(\mathcal{B}_i) - \mu(\mathcal{B}_i)_{EIG}}{\mu(\mathcal{B}_i)} \rangle|$ 
(circles), and $|\langle (\frac{ \mu(\mathcal{B}_i) - K \mu(\mathcal{B}_i)
_{EIG}}{\mu(\mathcal{B}_i)} \rangle|$ (diamonds) versus $\epsilon$, in a
log-log graph. The average is performed over intervals $\mathcal{B}_i$ 
centered in consecutive points of a trajectory of length 5000 and
$\mu(\mathcal{B}_i)_{EIG}$ is calculated using the set of all UPOs of
period 29. The standard deviation bar is also shown in this figure. Whenever
$\frac{\beta}{\alpha} <0$, which indicates the presence of low-period UPO
inside the interval, $\mu(\mathcal{B}_i)_{UPO}^d$ connot be neglegted, and
therefore the approximation proposed in Eq. (\ref{correction}) cannot be
used. Furthermore, the calculation of the measure from Eq. (\ref{mu_EIG})
oscillates widely in intervals dominated by low-period UPOs, since one 
considers different sets of UPOs with varying periods. Consequently, the
averages calculated in this figure are restricted to intervals with
$\frac{\beta} {\alpha} > 0$. Additionally, $\tau^{min}_{UPO}$ is considered to
be approximately given by $\tau^{min}_{REC}$. As a result, in an average
sense, the presence of recurrent trajectories (circles) significantly affects
the correctness of the calculation of the measure from the eingenvalues of the
UPOs in Eq. (\ref{mu_EIG}) (squares). It is apparent from the figure that, on
average, the correction proposed in Eq. (\ref{correction}) for $\mu_{EIG}$
yields a value closer to the real measure $\mu$, as well as a lower standard
deviation. 
\\
\indent
In conclusion, finite size regions of chaotic attractors can generally be
classified in two categories, hyperbolic and nonhyperbolic ones. For the
former, the measure is almost completely supported by the UPOs inside the
region and the distribution of the returns close to a Poissonian, while for
the latter, the contribution of the recurrent trajectories not associated to
UPOs inside the region is significant and the return time distribution
deviates from a Poissonian. In this case an exact calculation of the measure
can only be performed from the first return times. This has strong
implications on experiments, as there only finite region can be monitored and
many systems are nonhyperbolic. 
\\
\indent
M. S. B. and S. K. acknowledge support by the Alexander von Humboldt 
fundation, a stay at IMPA, and discussions with V. Sidoravicius and
O. de Almeida. C. G. and partially also M. S. B. were financed by FAPESP.


\begin{thebibliography}{99}
\bibitem{periodlength}
In the examples that follow, the border of short and long orbits is of 
a period of about 20. However, the exact value is not of crucial importance.
\end{thebibliography}
\end{document}